%% file: main.tex
\documentclass[sigconf,nonacm]{acmart}
\pdfoutput=1

\usepackage{booktabs}
\usepackage{graphicx}
\graphicspath{{figures/}}

\begin{document}

\title{Held While Waiting: What LLM Agent Sessions Reserve on Serverless and Agent Runtimes}

\author{Mekhala Hithyshini Sreenivasa}
\authornote{The author is employed by Amazon Web Services, which provides two
of the platforms studied here (Amazon Bedrock AgentCore Runtime and AWS
Lambda). This work was done independently, outside that employment, using the
author's own accounts and funds. It uses no non-public Amazon information, and
the views expressed are the author's own and do not necessarily reflect those
of Amazon.}
\affiliation{%
  \institution{Amazon Web Services}
  \city{Seattle}
  \country{USA}}
\email{mekhal@amazon.com}

\begin{abstract}
LLM agents spend nearly all their time waiting for the model. On Amazon
Bedrock AgentCore, our agents waited a median of 2.7 seconds for each model
reply, then computed for an estimated 5 to 40 milliseconds. AgentCore does
not charge for CPU during the wait. But it reserves a separate environment, a
virtual machine with its own memory, for every conversation, and keeps it
reserved until the conversation is closed. So 1,000 conversations reserve
1,000 environments to do the work of 2 to 15 processor cores. A service runs
out of environments long before it runs out of processing power.

The shortage has two causes, and each has a fix. An environment stays
reserved after its conversation ends: in the customer-service benchmark we
ran, a conversation lasts 33 seconds on average, but its environment is held
for 15 minutes more unless the client closes it. And an environment serves
only one conversation, which is mostly waiting: in our simulation, 64
conversations sharing one environment need 59 times fewer environments, and
99 of every 100 replies are delayed by at most 40 milliseconds.

The milliseconds are an estimate because they are hard to measure: a stubbed
model, a timer around the model call, and steps run back to back each gave a
plausible number that was too low. We show how to avoid all three. Of the
two fixes, closing needs only the client. Sharing needs the platform, and of
the agent runtimes we surveyed only one allows it, in part. They keep
conversations apart because an agent runs tools, and sometimes code, chosen
by a model. The open problem is to keep them apart without reserving an
environment for each while it waits.
\end{abstract}

\maketitle

\section{Introduction}
\label{sec:intro}
\input{sections/01-intro}

\section{Background}
\label{sec:background}
\input{sections/02-background}

\section{Method}
\label{sec:method}
\input{sections/03-method}

\section{The cost of a step}
\label{sec:step}
\input{sections/04-step}

\section{Capacity, not cost}
\label{sec:capacity}
\input{sections/05-capacity}

\section{What platforms bill, hold and limit}
\label{sec:platforms}
\input{sections/06-platforms}

\section{Evaluating the remedy}
\label{sec:remedy}
\input{sections/07-remedy}

\section{Lessons and open problems}
\label{sec:lessons}
\input{sections/08-lessons}

\section{Threats to validity}
\label{sec:threats}
\input{sections/09-threats}

\section{Related work}
\label{sec:related}
\input{sections/10-related}

\section{Conclusion}
\label{sec:conclusion}
\input{sections/11-conclusion}

\bibliographystyle{ACM-Reference-Format}
\bibliography{refs}

\end{document}

%% file: sections/01-intro.tex
Serverless platforms charge only while code runs. LLM agents fit that model
poorly. An agent session spends almost all of its time waiting on a
model, yet it holds an execution environment while it waits. A session
proceeds in steps, each one model call plus the framework work around it. In
our deployment, a step waits a median of 2.7\,s on the model and computes for
an estimated 5 to 40 milliseconds.

One of the platforms built specifically for agents over the last two years,
Amazon Bedrock AgentCore Runtime, stops billing for CPU while a session waits
on I/O~\cite{aws-agentcore-pricing}. That addresses the cost of waiting,
which was small to begin with: at published prices we calculate compute to be
a few per cent of the model-token bill (Section~\ref{sec:capacity}). It does not, however, change what a waiting session holds. Agent
runtimes give each session a dedicated, isolated environment for its whole
life~\cite{aws-agentcore-sessions,azure-hosted-agents}, and count it against a
session quota whether it is busy or idle~\cite{aws-agentcore-quotas}. On these runtimes, what limits an agent workload is therefore less what it
pays for processor time than how many isolated environments it holds, and for
how long.

This paper characterises agent steps on serverless and agent runtimes, and
measures that capacity. It makes four contributions.

\textbf{Agent runtimes, characterised} (Sections~\ref{sec:step}, \ref{sec:capacity}
and~\ref{sec:platforms}). First, we observe 3,549 steps of LangGraph and Strands agents serving a
public benchmark, deployed on AgentCore and self-hosted. Second, we run our
own CPU benchmark inside AgentCore and AWS Lambda, on the platforms' own
processors. Of the time a deployed agent step takes, only 0.2 to 1.5 per cent is
computation; the rest is waiting on the model. A platform that gives each
session its own environment holds that environment through the wait. At any
moment, then, only 2 to 15 of every 1,000 held environments are using a
processor, so for each core of work actually done the platform holds roughly
70 to 600 environments. A direct measurement of up to 400 concurrent sessions
checks this calculation. A survey of nine platforms' public documentation (our own;
Section~\ref{sec:platforms}) shows where that ratio applies: the agent
runtimes of AWS and Microsoft dedicate an environment per session, and Google's shares one only
when the agent's code is asynchronous, while every runtime that shares
environments by default is a general-purpose one~\cite{aws-agentcore-sessions,azure-hosted-agents,gcp-agent-engine-pricing,gcp-cloudrun-concurrency,aws-lambda-webfunctions}.

\textbf{What isolation costs, and the cheapest fix} (Section~\ref{sec:remedy}).
In a simulation driven by the deployed workload, sharing 64 sessions per
environment holds 59 times fewer environments than isolating each one, for at
most 40\,ms of added latency per step at the 99th percentile, against a 2.7\,s
model call. That fix needs the platform to change; the next needs only the
client. A session left to the default 15-minute idle timeout, instead of being
closed when its conversation ends, keeps its environment for the whole
timeout, and the same traffic then holds 28 times more environments than when
sessions are closed promptly. The multiple is simply the time an unclosed session is held over the time a
closed one is, $933/33 = 28$ for our 33-second conversations, but it is the largest
factor a client controls.

\textbf{A measurement method, and three ways to get it wrong}
(Section~\ref{sec:method}). To measure a framework's CPU per step without paying for thousands of model
calls, and so that every run replays the same conversation with no seconds
of network wait around each millisecond of computation, the model must be
replaced with something fake. How it is replaced decides what is measured. Three mistakes each give a plausible but
wrong number. (1)~Replacing the model \emph{object} with a stub skips the
work of serialising the growing conversation, so framework CPU looks flat as
conversations grow; faking only the network socket keeps that work.
(2)~Timing ``the model call'' in a deployed agent counts the client's own
serialisation and parsing as model time; by our estimate that hid most of
the framework's CPU in our deployed data. (3)~Benchmarking steps back to
back understates CPU: on our desktop, a step run after a wait of deployed length cost 1.9 to 4.7
times as much. Measured with the socket mock, framework CPU per step is linear in
conversation size in both frameworks and on every processor we measured,
though its constants vary more than fivefold.

\textbf{Open problems for platform builders} (Section~\ref{sec:lessons}). The main
one is how to keep each session confined, as agent runtimes do today by
giving it an environment of its own, without that environment sitting
reserved while the session waits on the model. Today's platforms offer one or the
other, at least among the nine we surveyed: the agent runtimes confine each
session in its own environment, and the runtimes that share environments
confine per tenant rather than per session. Isolates and per-tool sandboxes
point toward a middle ground (Section~\ref{sec:lessons}).

Two scoping statements frame everything that follows. First, the multiples we
report count what the \emph{customer} holds and pays for: environments, quota
slots and billed memory. They are not the provider's cost. A provider can
pack many idle microVMs onto one machine, so a thousand held environments
cost it far less than a thousand machines' worth; by how much, we cannot see
from outside. Our claims are about the customer's side. Second, we do not
argue that per-session isolation is a mistake. Agents run tools, and sometimes
code, chosen by a model on a user's behalf, and confining each session has real
value. We measure its price.

%% file: sections/02-background.tex
\textbf{Terms.} An LLM agent runs a loop. Each pass around it is a
\emph{step}: the agent calls a model with the conversation so far, and the
model either answers or requests one or more tool calls, which are executed
and appended to the conversation. A \emph{turn} is one run of the loop, from
a user's message until the agent replies; it usually takes several steps.
A \emph{session} is the sequence of turns that share one
conversation's state. A step's wall-clock time divides into time
\emph{blocked} on a remote service, the model or a tool, and time spent
\emph{computing} in the framework that runs the loop. This paper measures
both, per step, and asks what a hosting platform reserves while the first
dominates.

\textbf{What function platforms assume.} Function-as-a-service was designed
for short, largely CPU-bound request handling multiplexed across tenants.
Several design decisions follow from that, and each is sound for that
workload~\cite{hellerstein2019onestep,jonas2019berkeleyview}. One request
occupies one execution environment, which makes isolation simple and resource
use easy to attribute~\cite{aws-lambda-concurrency}. Billing is per
millisecond of the environment's lifetime, on the assumption that an
environment that is held is being used~\cite{aws-lambda-pricing}. And
capacity is limited by counting concurrent environments. All three treat an
invocation as a unit of work. For an agent step, an invocation is mostly a
unit of waiting.

\textbf{What agent runtimes changed.} Managed agent runtimes, introduced by
the major cloud providers over the last two years, adapt this model to agents
in two ways. First, the unit of reservation becomes the session: a client-chosen
session identifier pins every request of a conversation to the same
environment, which is kept until the client closes it, an idle timeout passes,
or a maximum lifetime is reached~\cite{aws-agentcore-sessions,azure-hosted-agents}.
Second, billing distinguishes active from blocked time: AgentCore, for example,
stops charging for CPU while a session waits on I/O~\cite{aws-agentcore-pricing}.
Section~\ref{sec:platforms} examines what these changes do and do not alter.
The first lengthens the reservation from one request to a whole conversation.
The second removes most of the bill for waiting but not the reservation
itself.

\textbf{Prior work, and the gap.} That serverless platforms reserve capacity
for functions blocked on I/O is established~\cite{stojkovic2023mxfaas,liu2025concurrency}.
What has not been measured is the agent case on the platforms built for it:
how much of an agent step is waiting, what a per-session runtime holds during
that wait, and what a customer can do about it
(Section~\ref{sec:related} reviews the related work).

%% file: sections/03-method.tex
The capacity results rest on two numbers per step: how long the
step waits on the model, which we call $w$, and how much processor time the
framework spends around that call, $c$. Section~\ref{sec:capacity} divides
one by the other. The two differ by two to three orders of magnitude, a wait
being seconds and the computation milliseconds. Measured together, with a
real model on the network, the milliseconds would be lost in the variation of
the seconds, so we measure each with an instrument built for it
(Table~\ref{tab:instruments}) and then check the combination against agents
running in a real deployment. This section describes the instruments and the
reason for each design choice.

\begin{table}[t]
\caption{The five instruments. The first four isolate one quantity each,
with the network removed to measure CPU and present to measure wait; the
fifth observes both in a deployment, but not cleanly
(Section~\ref{sec:method:deployed}).}
\label{tab:instruments}
\small
\begin{tabular}{@{}p{0.24\columnwidth}p{0.08\columnwidth}p{0.34\columnwidth}p{0.22\columnwidth}@{}}
\toprule
\textbf{Instrument} & \textbf{Gives} & \textbf{How} & \textbf{Scale} \\
\midrule
Transport mock, LangGraph & $c$ & socket faked; Apple M5 and Intel x86 & 4,800 steps per machine \\
Transport mock, Strands & $c$ & socket faked; Intel x86 & 4,800 steps \\
Transport mock, on platform & $c$ & socket faked; both frameworks, inside AgentCore and Lambda & 4,800 steps per configuration \\
Calibrated latency & $w$ & real calls to Bedrock and two HTTP endpoints & 50 model, 30 tool calls \\
Deployed agents & $w$, and part of $c$ & real agents on a public benchmark & 3,549 steps, 384 sessions \\
\bottomrule
\end{tabular}
\end{table}

\subsection{Framework CPU per step: faking the network, not the model}
\label{sec:method:mock}

Measuring what a framework does around a model call requires running the
agent without a real model, for three reasons. Cost: each 40-step run makes
41 model calls, and we make over 2,500 runs across machines and platforms.
Repeatability: a real model's replies vary from run to run, so no two runs
would carry the same conversation, and the request sizes we regress against
would differ. Isolation: with a real model, each millisecond of framework
work sits between seconds of network wait, and the two must be separated
before the former can be measured at all.

How the model is removed decides what is measured. Our first attempt
replaced the model \emph{object} inside the framework with a stub that returns
scripted tool calls, and CPU per step appeared flat as the conversation grew:
the stub skips exactly the work that grows, converting the history to the
provider's wire format, serialising and encoding the request, and parsing the
reply back into message objects.

We therefore keep the model client, so that the serialisation work the stub
skipped is executed and counted, and fake only the layer beneath it, the
network. Both frameworks reach Amazon Bedrock through the AWS SDK for Python, which lets a program
register a handler on its \texttt{before-send} event. The SDK calls the handler
``when the operation has been fully serialized, signed, and is ready to be sent
over the wire'', and if the handler returns a response, that response
``fulfills the request''~\cite{boto3-events}. Our handler returns a scripted
reply in Bedrock's own wire format (a tool call for 40 steps, then a final
answer), so every byte of request building, signing and response parsing
executes, and only the socket is absent. Before each sweep the harness checks
the one thing this design depends on, that the serialised request grows as the
conversation grows, and aborts otherwise. It grew from 1.1\,KB at the first
step to 19.6\,KB at the fortieth in every run.

\textbf{Two frameworks.} We measure LangGraph~\cite{langgraph} (version
0.6.11), which calls Bedrock through LangChain's \texttt{ChatBedrockConverse}
client, and Strands~\cite{strands} (1.55.0), AWS's agent SDK, which ships its
own Bedrock client. We chose them because they take opposite positions on who
owns the agent loop (LangGraph lets the developer define it; Strands runs it)
and because they are the two frameworks of the deployed study in
Section~\ref{sec:method:deployed}. To make their measurements comparable, both
agents register the same three tools with the same names, argument schemas
and return values, use no system prompt, and keep the whole conversation.
Strands by default keeps only a sliding window of recent messages, which
would stop the request from growing; we replace that manager with one that
keeps everything. Each run drives the agent through 40 steps of one model call
and one tool call. Forty steps take the request to 19.6\,KB (19,617 bytes
for LangGraph, 19,627 for Strands, on every machine and platform), the low end
of the sizes the deployed agents send (Section~\ref{sec:method:deployed}).
With payloads this close, a difference in CPU is a difference in code.

\textbf{Two clocks.} CPU time is what the operating system charges to a
thread or a process while it is running on a processor; time spent blocked is
not charged. We read it through Python's \texttt{time.thread\_time} and
\texttt{time.process\_time}, which return ``the sum of the system and user CPU
time'' of the current thread or process and do ``not include time elapsed
during sleep''~\cite{python-time}. Which clock to read depends on where a
framework puts its work.
LangGraph runs the model path on the calling thread but dispatches tools
through a pool of worker threads, so we read per-thread CPU and add the
worker's CPU explicitly. Strands runs the SDK call on a worker thread and
processes the reply on its event-loop thread, splitting one model call across
two threads (its Bedrock model class hands the SDK call to a worker thread and posts
each reply event back to the loop~\cite{strands-bedrock-src}), so per-thread
CPU would miss part of it; we instead read
process-wide CPU over the interval between Strands' own before- and
after-model-call hooks. Its loop is strictly sequential, so nothing else runs
in that interval. All runs are synchronous, because under Python's
\texttt{asyncio} a coroutine that suspends hands its thread to another, and
per-thread attribution stops meaning anything.

\textbf{Design of a sweep.} A sweep is 120 runs of 40 steps, 4,800 steps, on one
machine or platform configuration. The repeated runs average out scheduler and allocator noise, and the
per-step medians across them are what Section~\ref{sec:step} fits. For
LangGraph the 120 runs are split, 40 each, across three configurations of its
\emph{checkpointer}, the component that persists the graph's state after each
step so that a run can be resumed later: none, in-memory, and SQLite. The
three form a ladder that adds one cost at a time, separating the framework's
own cost from the cost of persisting state. Strands has no checkpointer as
such; its counterpart is an optional session manager that saves each message
as it is added~\cite{strands-session}, and an agent built without one, the
default, persists nothing. We measured Strands in that default only, which
matches LangGraph's \emph{none} configuration, so all 120 of its runs are in one
configuration.

\textbf{Machines and platforms.} Constants measured on one machine say little
about another, so we measure on several, and inside the platforms the paper
is about. LangGraph ran on an Apple M5 laptop (CPython 3.9.6) and both
frameworks on an Intel Core i5-8500T desktop (CPython 3.14.4). We then ran
the same transport-mock sweep inside Amazon Bedrock AgentCore Runtime, deployed as a code
package~\cite{aws-agentcore-code-deploy} into the microVMs the runtime
provisions per session~\cite{aws-agentcore-sessions}, and inside AWS Lambda on
both of its processor architectures, arm64 and x86-64. Lambda allocates CPU
in proportion to memory, with a full vCPU at 1,769\,MB~\cite{aws-lambda-memory},
so on Lambda we also swept memory (512, 1,024, 1,769 and 3,008\,MB) to see
whether a smaller function changes the CPU a step uses or only how long it
takes. On both platforms we forced a fresh environment for each of four groups
of 30 replicates, so that the results span more than one host, and we recorded
each host's processor; the platforms placed us on several. Library versions
are pinned identically everywhere. The interpreter is not (CPython 3.12 on
the platforms), so differences in constants between machines include the
interpreter's. The
socket is faked there too, so the cloud runs cost a few dollars and no tokens.

\subsection{Blocked time: calibrated latency}
\label{sec:method:latency}

The transport mock of Section~\ref{sec:method:mock} measures CPU with the
network removed, so the wait it removes must be measured separately. We
recorded 50 model calls to Amazon Bedrock (Anthropic's Claude Haiku~4.5,
through Bedrock's Converse API, in the us-east-1 region) on 3 September 2026,
from a client about 200\,ms of round trip from the endpoint. Model latency is
governed by the number of tokens generated rather than
consumed~\cite{kwon2023pagedattention,zhong2024distserve}, so we drew each
call's \texttt{max\_tokens} log-uniformly between 48 and 1,536, which gives
short replies as much weight as long ones, and recorded the realised output
length with each latency. We also recorded 30 calls to two public HTTP
endpoints as stand-in tools; the only property of a tool call that matters
here is that it blocks. A warm-up call to each endpoint was discarded, to
exclude connection and TLS set-up. The medians are 4.23\,s per model call and
0.97\,s per tool call. These waits define the laboratory scenario in
Section~\ref{sec:capacity}; Section~\ref{sec:step} shows that the deployed
agents' waits are shorter.

\subsection{A deployed anchor, and what it cannot see}
\label{sec:method:deployed}

The instruments above measure $c$ and $w$ separately: $c$ on a laptop, a
desktop, and inside AgentCore and Lambda, always with the model faked; $w$ on
real calls with no agent around them. Section~\ref{sec:capacity} combines them by calculation. To check that
calculation,
we use a dataset collected in September 2026 for a separate study: the same
two frameworks serving $\tau^2$-bench~\cite{barres2025tau2}, a public benchmark in
which a simulated user and an agent complete retail customer-service tasks
through tools, against Claude Sonnet~4.5, deployed three ways. LangGraph ran
in a self-hosted container and on AgentCore Runtime; Strands ran on AgentCore
Runtime. The dataset holds 384 sessions of 7 to 19 steps. We use its 3,549
\emph{warm} steps, those served by an already-running process, because a cold
step's server time includes process start-up, a different quantity from the
per-step cost measured here.

Each deployed step records server time, model time and their difference,
which that study called overhead. One property of this split matters. Model time was measured around the model-client call: around
\texttt{litellm.completion} for LangGraph, whose deployed agent reached
Bedrock through LiteLLM~\cite{litellm}, a library that translates between
model APIs, and between Strands' model-call hooks. That interval contains the
client's own request serialisation and response parsing, which is the work
Section~\ref{sec:method:mock} measures as framework CPU. The deployed overhead
therefore covers only the framework and server work \emph{outside} the model
client. It omits client serialisation, and it is wall time rather than CPU.
It bounds deployed CPU in neither direction.

We therefore bracket deployed CPU per step rather than claim to measure it.
The low end is the measured overhead outside the client. The high end adds the
client serialisation that the transport-mock fit for the same framework,
measured on the same platform, predicts at the deployed request size: a
deployed request carries a median of 6.9k input tokens, which at an assumed 3
to 5 bytes per token is 20 to 34\,KB. Section~\ref{sec:step} reports the
bracket. The lesson is that timing
``the model call'' quietly hides the client's own work.

%% file: sections/04-step.tex
Section~\ref{sec:method} built the instruments; this section reports what
they read: the wait $w$, the framework CPU $c$ on each machine and platform,
the checkpointer's extra CPU, and, since deployed CPU per step cannot be
measured directly, a low and a high estimate of it that
Section~\ref{sec:capacity} rests on.

\subsection{Wait tracks the answer's length, not the conversation's}

How long an agent waits for the model depends on how long the model's answer
is, not on how long the conversation has grown. On every step the agent
sends the whole conversation so far as \emph{input}, 5k to 10k tokens in our
deployment, and gets back a short \emph{output}, the answer: a median of 99
to 113 tokens. A model call waited a median of 2.74\,s for LangGraph and
2.59\,s for Strands, and 90\% of the 3,549 steps waited between 1.7 and
6.5\,s.

If the wait grew with the input, a conversation would wait longer at every
step. It does not. Regressing each step's wait on both token counts,
separately for the three deployed configurations
(Section~\ref{sec:method:deployed}), each extra output token adds 19.3 to
19.5\,ms, while extra input adds nothing ($-15$ to $-51$\,ms per thousand
tokens). Model servers explain why: they read the whole input in one
parallel pass (\emph{prefill}), then write the answer one token at a time
(\emph{decode}), so the slow part, and with it the wait, scales with the
answer~\cite{kwon2023pagedattention,zhong2024distserve}. Our deployment is consistent with this over the 5k to 10k inputs it covers.

Our calibration calls (Section~\ref{sec:method:latency}) waited longer, a
median of 4.23\,s, though their answers were about as long (median 137
tokens). The same regression places the gap in a fixed delay per call:
about 3.5\,s, against 0.9 to 1.2\,s in deployment. The two sets differ in
model (Haiku~4.5 against Sonnet~4.5), client location and date, and 50 calls
cannot tell these causes apart. We therefore use the deployed waits wherever
a realistic value is needed, and keep the calibrated ones as the laboratory
scenario.

\subsection{Framework CPU grows linearly with the request}
\label{sec:step:cpu}

\begin{figure}[t]
\centering
\includegraphics[width=\columnwidth]{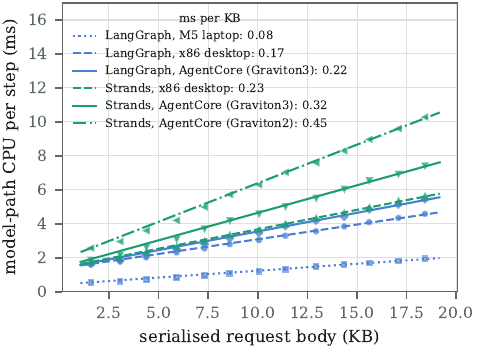}
\caption{Model-path CPU per step against the size of the request it
serialises, under the transport mock. Each point is the median across runs at
one of the 40 step positions; each line is the least-squares fit of
Table~\ref{tab:fits}. Blue is LangGraph and green Strands; line style marks
where the code ran: laptop, desktop, or inside AgentCore on two processor
generations. Every series is linear; no two share constants.}
\label{fig:cpu}
\end{figure}

The transport mock splits each step's CPU into two parts. The \emph{model
path} is everything the framework does around the model call: converting
the message history to the provider's format, serialising and signing the
request, parsing the reply, and updating its own state. \emph{Tool dispatch}
is everything it does around the tool call: validating the arguments the
model produced, running the tool, and appending the result.
Figure~\ref{fig:cpu} plots the model path against the size of the request it
serialises, for every framework and machine we measured off Lambda.

We model the model path as a fixed cost per step plus a cost proportional
to the request,
\[
\mathit{cpu} = c_{\mathrm{fixed}} + c_{\mathrm{KB}} \cdot \mathit{KB},
\]
and fit it by least squares through the median at each of the 40 step
positions (Table~\ref{tab:fits}). The line fits every configuration closely ($R^2$ of 0.96 to 0.98). It also
holds for single steps, which are noisy: fitted through all 4,800 individual
M5 steps, it gives the same slope.

\begin{table}[t]
\caption{Model-path CPU fits under the transport mock: CPU per step
$= c_{\mathrm{fixed}} + c_{\mathrm{KB}}\cdot\mathit{KB}$, over requests of
1.1 to 19.6\,KB. AgentCore rows ran inside the runtime's own microVMs, on the
Graviton3 (G3) or Graviton2 (G2) processors it placed us on. The last column
is tool dispatch, which is per step and does not grow.}
\label{tab:fits}
\small
\begin{tabular}{@{}lrrrr@{}}
\toprule
\textbf{Framework, machine} & $c_{\mathrm{fixed}}$ & $c_{\mathrm{KB}}$ & $R^2$ & \textbf{Tool} \\
 & (ms) & (ms/KB) & & (ms) \\
\midrule
LangGraph, Apple M5 & 0.43 & 0.081 & 0.96 & 0.22 \\
LangGraph, Intel x86 & 1.36 & 0.173 & 0.97 & 0.78 \\
LangGraph, AgentCore G3 & 1.35 & 0.220 & 0.96 & 0.73 \\
Strands, Intel x86 & 1.39 & 0.229 & 0.98 & 0.29 \\
Strands, AgentCore G3 & 1.41 & 0.324 & 0.98 & 0.30 \\
Strands, AgentCore G2 & 1.85 & 0.454 & 0.98 & 0.39 \\
\bottomrule
\end{tabular}
\end{table}

\textbf{The form transfers; the constants do not.} Every machine gives a
straight line, but each gives its own constants. The same LangGraph code
costs twice as much per KB on the x86 desktop as on the M5, and nearly three
times as much inside AgentCore. Even one runtime has no single set:
AgentCore placed our Strands sessions on two processor generations, and the
older one cost 1.4 times as much per KB. On the same desktop, the two
frameworks spend in different places. Strands' model path is steeper, while
LangGraph's tool dispatch costs nearly three times as much, because it
checks the model's arguments with Pydantic, a data-validation library, and
runs each call on a thread pool. Tool dispatch does not grow over the 40
steps in either framework: it handles one call, whatever came before. The
model path grows because a framework re-serialises the whole conversation on
every step.

\textbf{Inside Lambda, memory size changes how fast a step computes, not how
much.} We repeated the transport-mock benchmark in AWS Lambda at four memory
sizes on both processor architectures, 16 configurations in all
(Table~\ref{tab:lambda}). In every one the model path again grows in a
straight line with the request ($R^2$ 0.95 to 0.98), and its CPU per step
hardly changes with memory. What memory changes is speed. Lambda gives a
function a share of a processor in proportion to its memory, reaching one
full vCPU (virtual CPU: one processor core, as a cloud platform allocates
it) at 1,769\,MB~\cite{aws-lambda-memory}. At 512\,MB the same work
therefore took 3.2 to 3.6 times as long. Above 1,769\,MB it took no less,
because the agent loop runs on one thread, and a thread can use at most one
vCPU. A small function thus holds its environment a few times longer while
it computes, but that is still tens of milliseconds against a wait of
seconds. Lambda also placed us on mixed hardware: three Graviton generations
on arm64, and Xeons of two clock speeds on x86.

\begin{table}[t]
\caption{Model-path CPU per step on AWS Lambda, as fitted at a 19.6\,KB
request (ms), and the client-measured wall time of a group of 30 runs
relative to 1,769\,MB (one vCPU), by memory size.}
\label{tab:lambda}
\small
\begin{tabular}{@{}rrrrrr@{}}
\toprule
\textbf{Memory} & \multicolumn{2}{c}{\textbf{arm64}} & \multicolumn{2}{c}{\textbf{x86-64}} & \textbf{Wall} \\
(MB) & LangGraph & Strands & LangGraph & Strands & \textbf{time} \\
\midrule
512 & 7.0 & 10.1 & 7.6 & 10.8 & 3.2--3.6$\times$ \\
1,024 & 6.7 & 9.6 & 6.0 & 10.4 & 1.6--1.7$\times$ \\
1,769 & 6.8 & 9.4 & 6.3 & 9.6 & 1$\times$ \\
3,008 & 6.8 & 7.6 & 6.7 & 9.6 & 0.8--1.0$\times$ \\
\bottomrule
\end{tabular}
\end{table}

\subsection{Checkpointing serialises the same history a second time}

A checkpointer (Section~\ref{sec:method:mock}) saves the graph's state after
each step. LangGraph keeps that state in named \emph{channels}, one of which
is the message history, and it already saves incrementally: a saver is
handed the new checkpoint together with ``new channel versions as of this
write'', and the in-memory saver stores a serialised copy of exactly those
channels~\cite{langgraph-checkpoint-src}. In an agent loop, though, the
channel that changes on every step is the message history, an append-only
log held in one channel. It has a new version every step, so every step
serialises the whole of it again, and the step serialises the same growing
object twice: once into the model request, once into the checkpoint. The
in-memory checkpointer's CPU per step (median across runs) grows from under 0.1\,ms at the first step to
0.56\,ms at the fortieth on the M5, and from 0.2 to 0.99\,ms on the
desktop. The SQLite checkpointer shows a per-step median of zero with spikes of up to
1.8\,ms, the pattern of batched writes; we did not measure where the flushed
work is charged, so report no total for it. On a serverless host
the process does not survive between invocations, so production agents
checkpoint to a remote store, and each write adds a network round trip:
blocked time, the quantity this paper is about. Our checkpoint figures are
therefore a lower bound.

\subsection{Deployed CPU per step, bracketed}

As Section~\ref{sec:method:deployed} explained, the deployed data cannot give
CPU per step directly. The deployed agents spent a median of 4.48\,ms (LangGraph on AgentCore) and
4.49\,ms (Strands) per step outside the model client. For LangGraph that
overhead also grows with the conversation, at about 0.25\,ms per thousand
input tokens ($R^2 = 0.99$ over medians binned by thousand tokens). For
Strands it does not, because Strands performs its serialisation inside the
model-call window, which the deployed overhead excludes.

We therefore bracket. The low end is the measured overhead, 4.5\,ms. The
high end adds the client serialisation that the AgentCore fits predict at
the deployed request size, 20 to 34\,KB. For LangGraph that is the AgentCore
row of Table~\ref{tab:fits}; for Strands it is one fit through the runs on
both processor generations together (1.73\,ms $+$ 0.41\,ms/KB). This gives
4.5 to 13.2\,ms for LangGraph and 4.5
to 20.3\,ms for Strands. The high end overstates the CPU of steps run back to back, for two reasons:
the overhead is wall time, which is at least the CPU inside it, and for
LangGraph the message conversion that the overhead already contains is
counted again by the fit. Against the 2.6 to 2.7\,s median wait, computation
is between 0.16\% ($4.5/2{,}740$) and 0.8\% ($20.3/2{,}590$) of the time a
deployed step takes. Section~\ref{sec:capacity} starts from this bracket and
then finds, by direct measurement, that a step costs more CPU after a wait of
deployed length than back to back, which raises the upper end to about 1.5\%.

%% file: sections/05-capacity.tex
\begin{figure*}[t]
\centering
\includegraphics[width=0.8\textwidth]{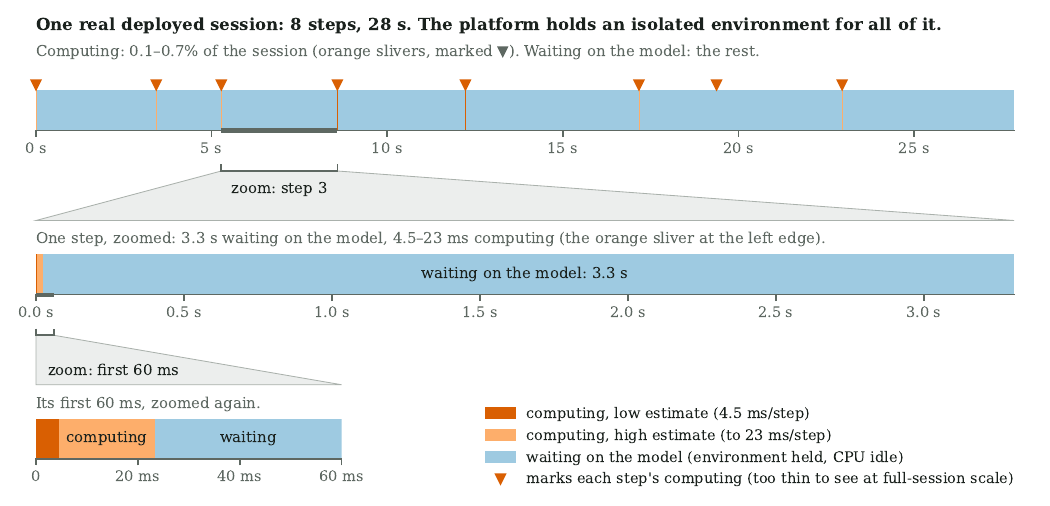}
\caption{One real deployed LangGraph session on AgentCore Runtime, drawn to
scale: the session closest to the median in both step count and total model
wait. Blue is time spent waiting on the model, during which the session's
environment is held and its CPU idle; orange is framework computation, drawn
at our low and high estimates of CPU per step (Section~\ref{sec:step}), and
marked with a triangle in the top row, where it is too thin to see. The
lower rows zoom into the third step and then into its first 60\,ms, where the
computation first becomes visible.}
\label{fig:timeline}
\end{figure*}

\textbf{The quantity that matters.} Section~\ref{sec:step} gave a step's
two durations, the wait $w$ and the CPU $c$; this section turns them into
capacity. Consider a platform that dedicates an environment to each session
in progress. Each environment is then on a processor
for a fraction $c/(w+c)$ of the time, so by Little's
law~\cite{little1961proof} $N$ held environments keep $N \cdot c/(w+c)$
processor cores busy on average. Equivalently, the platform holds
\[
\frac{w + c}{c}
\]
environments for every core of work it performs. This ratio is a property
of the workload: it does not depend on the quota or the arrival rate.
Figure~\ref{fig:timeline} shows why it is so large for agents. One deployed
session, drawn to scale, holds its environment for 28 seconds and computes
for at most 0.7\% of them.

\textbf{Our deployed agents hold tens to hundreds of environments per core
of work.} We apply the formula with each framework's median deployed wait
$w$ from Section~\ref{sec:step}: 2.74\,s for LangGraph and 2.59\,s for
Strands. Section~\ref{sec:step} could only bracket the CPU per step $c$, so
we apply it twice, once at the low estimate and once at the high one. For
LangGraph at the low estimate, $c = 4.5$\,ms, the platform holds
$(2{,}740 + 4.5)/4.5 \approx 610$ environments per busy core. Put the other
way round, 1,000 held sessions keep $1{,}000 \times 4.5/2{,}744.5 \approx 1.6$
cores busy. Table~\ref{tab:range} gives all four cases: roughly 70 to 600
environments per busy core.

\begin{table}[t]
\caption{Environments held per busy core. The first four rows are calculated
from the formula with the median deployed wait; the last two are measured
directly, with waits drawn from the deployed calls.}
\label{tab:range}
\small
\begin{tabular}{@{}lrrrr@{}}
\toprule
\textbf{Framework,} & $w$ & $c$ & \textbf{Busy cores per} & \textbf{Environments} \\
\textbf{CPU} & (s) & (ms) & \textbf{1,000 sessions} & \textbf{per busy core} \\
\midrule
LangGraph, low & 2.74 & 4.5 & 1.6 & 610 \\
LangGraph, high & 2.74 & 23.4 & 8.5 & 120 \\
Strands, low & 2.59 & 4.5 & 1.7 & 580 \\
Strands, high & 2.59 & 38.9 & 15 & 68 \\
\midrule
LangGraph, measured & drawn & 13--17 & 3.8--5.0 & 200--260 \\
Strands, measured & drawn & 16--31 & 4.8--9.3 & 110--210 \\
\bottomrule
\end{tabular}
\end{table}

The low estimate is the overhead measured outside the model client. The
high estimate adds the client serialisation that the fit of
Table~\ref{tab:fits} predicts, multiplied by 2.2. That factor is the extra
CPU a step costs after a real wait: one LangGraph session alone in its
process, with deployed waits, cost 14.8\,ms per step, against 6.8\,ms back
to back ($14.8/6.8$; measured below). This gives
$4.5 + (14.8/6.8) \times (13.2 - 4.5) = 23.4$\,ms for LangGraph and
$4.5 + (14.8/6.8) \times (20.3 - 4.5) = 38.9$\,ms for Strands. The two
frameworks span almost the same range, so its width is our uncertainty about
deployed CPU, not a difference between them. Two smaller points. Because the
waits are right-skewed, the mean wait, 3.3\,s, would give about 1.2 times
more environments per core than the median. And the laboratory benchmark,
with less CPU per step and the longer calibrated waits, keeps only 0.14 to
0.45 cores busy per 1,000 sessions: thousands of environments per core.

\textbf{Checking the calculation by direct measurement.} The numbers above
are calculated. To check them, we also measured busy cores directly. On the
x86 desktop we ran 50, 100, 200 and 400 sessions at once, spread over five
processes. Each session was the transport-mocked agent of
Section~\ref{sec:method:mock}, and every model call blocked for a wait drawn
from that framework's deployed calls. Over five-minute windows, the CPU that
the processes consumed grew with the number of sessions. Scaled to 1,000 sessions, LangGraph kept 3.8 to 5.0 cores busy and
Strands 4.8 to 9.3 (the last two rows of Table~\ref{tab:range}): 200 to 260
and 110 to 210 environments per busy core, both inside the calculated
ranges.

The direct measurement also shows more CPU per step than the same agents
run back to back in the same harness: 13 to 17\,ms for a LangGraph step
instead of 6.8\,ms (1.9 to 2.5 times), and 16 to 31\,ms for a Strands step
instead of 6.4\,ms (2.4 to 4.7 times, falling as sessions were added).
Concurrency is not the cause: one session alone in its process, with the
same waits, cost 14.8\,ms (LangGraph) and 27\,ms (Strands) per step. Nor is
garbage collection: with the collector disabled, LangGraph still cost
16.5\,ms. That leaves the wait itself: on our desktop, a step that ran after seconds
of idleness cost 1.9 to 4.7 times the CPU of the same step run back to back.
We did not separate the causes. The likely ones are a processor that has
lowered its clock and caches that have gone cold, costs every woken thread
pays; Strands' model call crosses two threads (Section~\ref{sec:method:mock}),
which would explain its larger penalty. The deployed overhead of
Section~\ref{sec:step} was measured in place and already includes this
effect. The serialisation predicted by the fit does not, which is why the
high estimates multiply that part by 2.2, LangGraph's factor. We use it for
Strands too.
Strands' own factor is larger ($27/6.4 \approx 4.2$) and would push its high
estimate higher still, but there is no need: measured directly, a Strands
step costs at most 31\,ms, already below the 38.9\,ms high estimate.

\textbf{What these numbers count.} They count what a customer holds
and plans against: environments, quota and billed memory. They do not count
what the provider spends: a provider can oversubscribe idle
microVMs~\cite{agache2020firecracker}, so an idle session costs it less, by
an amount we cannot observe.

\textbf{Why compute cost is the wrong lens.} An agent step incurs two
bills: one for the model's tokens, and one for the compute that hosts the
agent, waiting included. At published prices, we calculate the compute bill
to be a few per cent of the token bill where wall-clock time is
billed~\cite{aws-lambda-pricing}, and less where waiting CPU is not. So cost
is a weak argument against serverless agents. Not billing waiting CPU, as AgentCore does, lowers that bill further but
does not reduce the environments held.
Nor does a higher session quota, which moves the limit without changing the
ratio; the quota is also account-wide, shared with every other workload.

\textbf{The ratio narrows as context grows.} Model wait does not grow with
context (Section~\ref{sec:step}), but framework CPU does, so longer
conversations hold fewer environments per busy core. Extrapolating the
AgentCore-measured CPU fit, a 50k-token context would hold about 26 to 47
environments per busy core, and a 250k-token context about 6 to 11, the lower
figures including the extra CPU a step costs after a wait. These are
extrapolations far outside the 5k to 10k tokens we observed, and they hold
the wait at its short-context median, which prefill over 250k tokens would
lengthen. The direction is the useful finding:
the mismatch between what is held and what is used is \emph{largest for short
conversations}.

%% file: sections/06-platforms.tex
Sections~\ref{sec:step} and~\ref{sec:capacity} measured what an agent step
does. This section asks what platforms do with it: what each bills while a
step waits, what it holds, and what it limits. We surveyed the public
documentation of nine platforms on 27 October 2026
(Table~\ref{tab:platforms}): the managed agent runtimes of AWS, Microsoft and
Google; AWS Lambda in its four current forms; and two runtimes long built on
sharing, Cloud Run and Cloudflare Workers. Several of these pages changed
during 2026.

\begin{table*}[t]
\caption{What nine platforms bill, hold and limit while an agent waits on a
model. From public documentation, accessed 27 October 2026. ``Blocked'' is
waiting on the model. ``One session'' is a dedicated environment for each
conversation; ``many'' is one environment serving concurrent requests.
Durable functions are a feature of Lambda functions, listed separately
because they hold differently.}
\label{tab:platforms}
\small
\begin{tabular}{@{}p{0.17\textwidth}p{0.22\textwidth}p{0.21\textwidth}p{0.20\textwidth}p{0.11\textwidth}@{}}
\toprule
\textbf{Platform} & \textbf{Billed while blocked} & \textbf{Held while blocked} & \textbf{Default limit} & \textbf{Per environment} \\
\midrule
\multicolumn{5}{@{}l}{\emph{Agent runtimes}} \\
AgentCore Runtime~\cite{aws-agentcore-pricing,aws-agentcore-sessions,aws-agentcore-quotas} & CPU: no. Memory: yes, for the whole session & a dedicated microVM & 5,000 active sessions (2,500 in most regions) & one session \\
Foundry hosted agents~\cite{azure-hosted-agents,azure-hosted-agents-limits} & CPU and memory of active sessions & a dedicated sandbox, until idle timeout (15\,min default) & 1,000--2,000 sessions per region & one session \\
Vertex AI Agent Engine~\cite{gcp-agent-engine-pricing} & allocated vCPU and memory; idle time between turns not billed & instances (1 to 100 by default) & 1,000 instances max & 9 by default; 1 for synchronous code \\
\midrule
\multicolumn{5}{@{}l}{\emph{Function platforms}} \\
Lambda functions~\cite{aws-lambda-concurrency,aws-lambda-pricing} & wall-clock memory-time & an environment per request & 1,000 concurrent per region & one request \\
Lambda MicroVMs~\cite{aws-lambda-microvms} & running: compute; suspended: storage & a MicroVM, running \emph{or suspended}, against a memory quota & 400\,GB memory per region & one session \\
Lambda durable functions~\cite{aws-lambda-faq} & no compute during waits; steps billed & waits release the slot; a model call inside a step does not & 1,000 concurrent & one request \\
\midrule
\multicolumn{5}{@{}l}{\emph{Multiplexed runtimes}} \\
Lambda web functions~\cite{aws-lambda-webfunctions} & active CPU & a shared environment & account vCPU quota & many \\
Lambda Managed Instances~\cite{aws-lambda-managed-instances} & instance-hours & the customer's instance fleet & instance quotas & many \\
Cloud Run~\cite{gcp-cloudrun-concurrency} & while $\geq$1 request in flight & a shared instance & 100 instances per region & 80 (up to 1,000) \\
Cloudflare Workers~\cite{cloudflare-workers-limits} & CPU time only & a shared isolate & no general limit & many \\
\bottomrule
\end{tabular}
\end{table*}

\textbf{AgentCore does not bill waiting CPU; the reservation remains.} AgentCore does
not charge for processor time while an agent waits: ``CPU scales to zero
during I/O wait (waiting for LLM responses, tool / API calls, or database
queries)''~\cite{aws-agentcore-pricing}. That addresses the cost of the
waiting, which Section~\ref{sec:capacity} found to be a small share of the
bill. It does not address what the waiting holds. First, the environment: each
AgentCore session receives a dedicated
microVM~\cite{aws-agentcore-sessions}, and our deployed data shows it kept
between steps: in all 256 AgentCore sessions, every step ran in the same
process, whose age rose steadily across idle gaps of up to 10\,s, and no
step after the first started cold. Second, memory is billed across the
session, ``from microVM ready, initialization, active processing, idle
periods, until session termination''; the pricing page's example charges
60 seconds of CPU and 600 seconds of memory for a ten-minute
session~\cite{aws-agentcore-pricing}. Third, every session,
idle or not, counts against an account quota of 5,000 active
sessions, 2,500 in most regions~\cite{aws-agentcore-quotas}. Microsoft's Foundry hosted agents hold
in the same way, ``each session gets a dedicated
sandbox''~\cite{azure-hosted-agents}, and Microsoft's guide to
serving several users from one session states that ``one session per user doesn't
scale''~\cite{azure-multiplex}.

\textbf{Suspension removes the cost of idle time, not its capacity.} Lambda
MicroVMs, which give each session a microVM of its own, can be suspended
when idle. A suspended MicroVM pays only for snapshot storage, but it still
counts against the memory quota, which covers MicroVMs ``in the RUNNING or
SUSPENDED state''~\cite{aws-lambda-microvms}. Suspension also happens only
after a configured period without traffic, never in the middle of a request,
so it cannot reclaim the time a step spends waiting on a model. Durable
functions, Lambda's feature for long-running workflows, do release their
concurrency slot during a wait~\cite{aws-lambda-faq}. The waits they mean,
though, are timers and callbacks; by our reading of the documentation, a
model call made inside a step holds the slot for its duration.

\textbf{Sharing is already offered, outside agent runtimes.} Letting one
environment serve many requests in progress removes the per-session
reservation. What usually limits sharing is contention for CPU, and a
workload that waits for more than 98\% of the time, as ours does
(Section~\ref{sec:capacity}), creates little. This observation is not new.
MXFaaS~\cite{stojkovic2023mxfaas} and later work~\cite{liu2025concurrency}
make the general point, Cloud Run has shipped it for
years~\cite{gcp-cloudrun-concurrency}, and in 2025 and 2026 AWS shipped it in
Lambda, through Managed Instances and web
functions~\cite{aws-lambda-managed-instances,aws-lambda-webfunctions}. The
pattern in Table~\ref{tab:platforms} is that \emph{every surveyed runtime
that shares an environment among requests is a general-purpose one, while
the agent runtimes isolate per session}. Vertex AI Agent Engine is the
partial exception: an instance accepts several concurrent requests, nine by
default, but only if the agent's code is
asynchronous~\cite{gcp-agent-engine-pricing}.

\textbf{Per-session isolation is a choice, and it has a reason.} An agent executes tools and
sometimes code chosen by a model, on behalf of one user. A dedicated microVM
per session confines whatever that session does.
Section~\ref{sec:remedy} asks what that choice costs at a fixed level of
traffic, and what sharing would recover.

%% file: sections/07-remedy.tex
Section~\ref{sec:platforms} found that the agent runtimes hold one
environment per session. This section puts numbers on that choice: how many
environments 1,000 concurrent sessions hold under isolation, and how many
they would hold if sessions shared. Every result here is simulated, from
measured inputs.

\textbf{Workload.} Rather than run thousands of agents, we simulate them,
replaying the deployed sessions of Section~\ref{sec:method:deployed} as a
sequence of events: a session arrives, a step waits on the model, the step
computes, the session ends (a discrete-event simulation). Each simulated
session takes its number of steps from one of the 384 deployed sessions
(median 9, range 7 to 19), and each step's wait from one of the 1,094
deployed LangGraph model calls (median 2.74\,s). CPU per step is fixed
within a run at one of four values: the low and high estimates of
Section~\ref{sec:step}, 4.5 and 20.3\,ms; LangGraph's midpoint between
them, 11.7\,ms; and the highest estimate of Section~\ref{sec:capacity},
38.9\,ms. New sessions arrive at random, independently of one another (a
Poisson process), at about 30 per second; since a session lasts 33\,s on
average, that keeps about 1,000 in progress at any moment ($30 \times
33$). Each run covers 40 minutes and discards the first 20, so that sessions
left to the 15-minute idle timeout (below) have time to build up before we
count. As a check, the simulated environment counts agree with Little's
law~\cite{little1961proof} to within 1\%.

\textbf{Policies.} Under \emph{isolation}, each session holds its own
environment, with a vCPU to itself, from its first step to its end. We model
two endings. Either the client closes the session when the conversation
ends, or it does not, and the platform reclaims the environment after its
idle timeout, 15 minutes by default on
AgentCore~\cite{aws-agentcore-sessions}. Under \emph{multiplexing}, an
environment of 2 vCPUs hosts up to $k$ sessions. A new session goes to the
first environment with a free slot, and a new environment starts only when none has one.
When more steps need CPU than an environment has vCPUs, they queue in
arrival order, so the cost of sharing appears as delay added to a step.

\begin{figure}[t]
\centering
\includegraphics[width=\columnwidth]{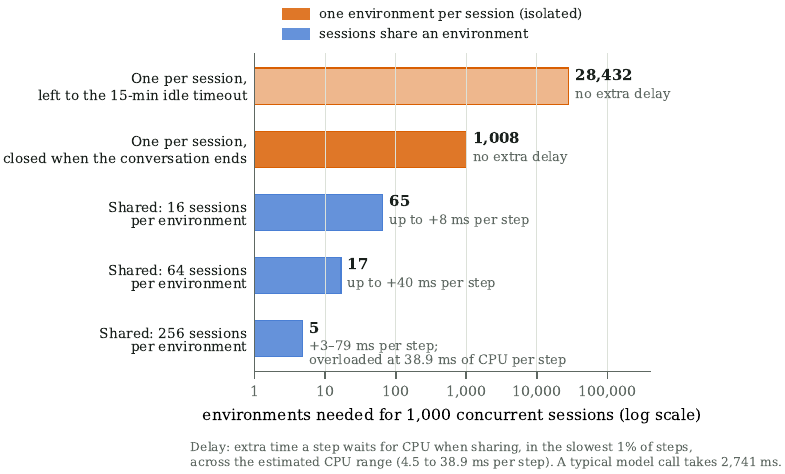}
\caption{Environments needed to serve 1,000 concurrent agent sessions under
each hosting choice, simulated with the deployed workload. Orange: one
environment per session, either closed when the conversation ends or left to
the platform's 15-minute idle timeout. Blue: sessions sharing environments of
2 vCPUs. Beside each bar is the extra delay sharing adds to a step, in the
slowest 1\% of steps, across those four CPU values.}
\label{fig:remedy}
\end{figure}

\textbf{Isolation holds an environment per session; sharing needs a few.}
Figure~\ref{fig:remedy} shows the result. With sessions closed promptly,
isolation holds 1,008 environments on average, one for each session in
progress.
Multiplexing 64 sessions per environment holds 17, which is 59 times fewer
($1{,}008/17$); at 256 per environment it holds 5. These counts are the same
at every CPU level, with one exception below. Memory depends on a footprint
we can only assume: at 0.5\,GB per environment plus 10\,MB per session, it
falls from 514\,GB to 18.5\,GB at $k = 64$.

\textbf{Sharing adds little delay up to $k = 64$.} The delay beside each bar is
what sharing adds to a step in the slowest 1\% of steps (the 99th
percentile, p99). At $k = 64$ it is 13\,ms at 20.3\,ms of CPU per step and 40\,ms at
38.9\,ms, at most 1.5 per cent of the 2.74\,s median model wait, with the
vCPUs at most 35\% busy. At $k = 256$ it is 79\,ms at 20.3\,ms of CPU,
with the vCPUs 63\% busy; at 38.9\,ms the sessions ask for more CPU than
the two vCPUs have, steps queue for seconds, and a sixth environment is
needed. Up to $k = 64$ the environments are far from saturated: as
Section~\ref{sec:capacity} found, there is very little computation to
share.

\textbf{Unclosed sessions hold 28 times as many environments.} The top bar of
Figure~\ref{fig:remedy} leaves sessions to the idle timeout instead. Each
then holds its environment for 933\,s rather than 33\,s, so the same traffic
holds 28,432 environments, 28 times as many ($933/33$). The fix is one call by the client. Against
AgentCore's largest default quota, 5,000 sessions, it is the difference between
about 150 new conversations per second ($5{,}000/33$) and about 5
($5{,}000/933$). AgentCore's newer runtime lowers the
\emph{memory} an idle session is billed for after 120\,s, but the session,
and its place in the quota, stays held until the
timeout~\cite{aws-agentcore-pricing,aws-agentcore-quotas}.

\textbf{What the simulation does not show.} It prices isolation; it does not
argue that isolation is unnecessary. Sessions sharing a process share its
memory and its failures, and an agent that runs tools or code chosen by a
model may need more confinement than that. The simulation also has no
think time between steps, as in an autonomous agent; with a human pausing
between turns, sessions would be longer and the idle timeout a smaller
multiple of them. It gives each isolated
session a whole vCPU, so it is generous to isolation. And it counts what the customer holds, environments and quota slots, not the
provider's physical cost (Section~\ref{sec:capacity}).

%% file: sections/08-lessons.tex
\subsection{For teams deploying agents}

\textbf{Close sessions explicitly.} On a per-session runtime, a conversation
that ends without its client closing the session keeps its environment until
the idle timeout. The multiple is 28 for our 33-second conversations against a 15-minute
default ($933/33$; Section~\ref{sec:remedy}). It is
the one factor a team can remove without changing platform.

\textbf{Size against quotas, not CPU.} In our deployment, 1,000 concurrent
sessions keep only 1.6 to 15 cores busy, by our calculation, yet hold 1,000
environments and 1,000 quota slots (Section~\ref{sec:capacity}). A plan that
watches CPU utilisation will therefore show plenty of headroom while the
session quota is already used up. Forecast concurrent sessions instead: the
arrival rate multiplied by how long each session is held, including any time
it sits idle before it is closed or times out.

\textbf{Choose isolation deliberately.} Where an agent's tools and code do
not require a sandbox per session, a multiplexed runtime, in our simulation,
holds 59 times fewer environments for the same traffic, at a latency cost of at most
40\,ms per step at the 99th percentile (Section~\ref{sec:remedy}). Where they do, the price
of isolation is a known quantity and can be budgeted.

\textbf{Benchmark with the socket faked, not the model.} Replacing the model
object with a stub removes exactly the framework work that grows with the
conversation (Section~\ref{sec:method:mock}). And when timing a model call in
a deployed agent, decide in advance whether the client's own serialisation
counts as model time or framework time. In our deployed data it was counted
as model time, which by our estimate hid most of the framework's CPU per step
(Section~\ref{sec:method:deployed}). Finally, measure CPU per step with
realistic waits between steps: a step that follows seconds of idleness cost
1.9 to 4.7 times the CPU of the same step benchmarked back to back, on our desktop (Section~\ref{sec:capacity}).

\subsection{Open problems}

Table~\ref{tab:transfer} sums up which established serverless techniques
carry over to agent workloads, and why. The problems below are the gaps it
leaves.

\textbf{Isolation without an environment per session.} Agent runtimes dedicate an
environment per session for isolation; multiplexed runtimes share one for
capacity. No platform in Table~\ref{tab:platforms} offers both. Candidate
mechanisms exist: per-session sandboxes inside a shared host process,
lightweight isolates (separate sandboxes within one
process, as in Cloudflare Workers), and snapshot-and-suspend that also releases the quota
slot. Which of them preserves the confinement agents need, at the density
their idleness allows, is open.

\textbf{A scheduling signal for blocked work.} Autoscalers observe request rate, queue depth and CPU utilisation, all
nearly uninformative for agents: a thousand blocked sessions and a thousand computing ones look
the same by request rate, and both look idle by CPU. What predicts the
resources an agent workload holds is the number of in-flight sessions blocked
on a remote call. None of the surveyed platforms documents making that number
available to its scheduler or its tenants.

\textbf{Quotas in the right unit.} Session quotas count environments. An
agent session's real demand is memory held over time, with almost no CPU.
A quota on memory-time, the memory a session holds multiplied by how
long it holds it, would charge each session for what it actually reserves,
rather than counting every session as one whole environment.

\textbf{Saving only what changed.} LangGraph saves an agent's state after
every step, but it treats the whole message history as a single item, so
each save writes out the entire conversation again, although only the latest
messages are new (Section~\ref{sec:step}). If the history were kept as a log
to which each step only appends its new messages, each save would stay small
however long the conversation grew. Shared-log serverless systems already
provide such logs~\cite{jia2021boki}. Whether agent frameworks can save
state this way without changing how agents are written is open.

\begin{table}[t]
\caption{Which serverless techniques transfer to agent workloads, in our
assessment.}
\label{tab:transfer}
\small
\begin{tabular}{@{}p{0.36\columnwidth}lp{0.36\columnwidth}@{}}
\toprule
\textbf{Technique} & \textbf{Transfers?} & \textbf{Why} \\
\midrule
Per-instance concurrency~\cite{stojkovic2023mxfaas,liu2025concurrency} & Yes & removes the reservation; agent runtimes do not use it \\
Lightweight isolation~\cite{agache2020firecracker,oakes2018sock,shillaker2020faasm} & Yes & density is the right goal when most sessions idle \\
Workflow orchestration~\cite{liu2023unum} & Yes & externalises control flow across long waits \\
Snapshot and restore~\cite{du2020catalyzer,cadden2020seuss,ao2022faasnap} & Partly & cuts start-up, not the wait \\
Stateful serverless~\cite{sreekanti2020cloudburst,jia2021boki,zhang2020beldi} & Partly & right abstraction, wrong granularity today \\
Suspension when idle & Partly & frees cost between turns, not the quota slot \\
Warm pools & No & pays to hold more idle environments \\
Autoscaling on request rate & No & blind to blocked versus active sessions \\
\bottomrule
\end{tabular}
\end{table}

%% file: sections/09-threats.tex
\textbf{Construct.} Deployed CPU per step is bracketed rather than measured
(Section~\ref{sec:method:deployed}); its upper end may
double-count work that its two instruments both include.

\textbf{Internal.} The transport mock removes the network but keeps every
other operation, which we verify by requiring the request to grow with the
conversation. It does not reproduce response sizes beyond the scripted ones,
so response-parsing cost is that of short replies. On AgentCore, health-check pings
arrive about every two seconds and fall in Strands' process-time windows;
each costs microseconds.

\textbf{External.} We deployed one workload, $\tau^2$-bench retail, whose tools are
executed by the benchmark rather than the agent, so deployed steps contain
no tool wait; an agent that runs its own tools is more blocked, not less. We
used one model family from one provider and two frameworks at pinned
versions, and measured contexts only up to 10k tokens. CPU constants vary more than fivefold across the six processor types we
measured, and within one platform (Section~\ref{sec:step}); only the linear
form is claimed to transfer. The extra CPU a step costs after an idle wait was measured on one desktop
machine, for both frameworks; we did not separate its causes, and did not measure
it on the cloud platforms, whose processors a hypervisor manages.

\textbf{Affiliation.} The author works for Amazon Web Services, whose
platforms are among those measured. All platform facts come from public
documentation as of 27 October 2026, and will change. All measurements ran on a personal
account at public prices.

%% file: sections/10-related.tex
\textbf{Serverless and blocked work.} That serverless platforms reserve
capacity for functions blocked on I/O is established. MXFaaS observes that
function execution is ``most of the time blocked on synchronous wait
operations''~\cite{stojkovic2023mxfaas}, and concurrency-informed
orchestration examines how concurrent requests should share
containers~\cite{liu2025concurrency}. Hellerstein et al.~\cite{hellerstein2019onestep}
and the Berkeley view~\cite{jonas2019berkeleyview} catalogued where serverless
fits data-intensive work poorly; we take these as given. Our contribution is
to measure the agent case, on platforms built for it.

\textbf{Serverless and LLMs.} Samanta and Nguyen study serverless LLM
\emph{inference} serving~\cite{samanta2025illuminating}; we study agent
\emph{orchestration}, where the unit is a multi-step session and the blocked
time is spent waiting on inference rather than performing it.

\textbf{Cold start and isolation.} Snapshot-restore~\cite{du2020catalyzer,cadden2020seuss,ao2022faasnap},
lightweight isolation~\cite{agache2020firecracker,oakes2018sock,shillaker2020faasm}
and the survey of cold-start work~\cite{golec2025coldstart} address start-up
cost and density. They bear on the open problem of Section~\ref{sec:lessons},
isolating sessions without giving each its own environment, more than on the
wait itself.

\textbf{Stateful serverless.} Cloudburst~\cite{sreekanti2020cloudburst},
Boki~\cite{jia2021boki} and Beldi~\cite{zhang2020beldi} supply durable state
and exactly-once semantics to functions; workflow systems such as
Unum~\cite{liu2023unum} externalise long-running control flow. None has been
applied to agent state, whose access pattern, an append-only conversation,
fits a shared log more naturally than the key-value model current
checkpointers assume.

%% file: sections/11-conclusion.tex
AgentCore, a managed agent runtime, does not bill CPU while a session waits.
Each session still holds an isolated environment, its memory
and a quota slot for its whole life. In our deployment, a session computes for between 0.2 and 1.5
per cent of that life, so the platform holds tens to hundreds of environments
for every core of work it performs. That is the price of isolating each session. This paper measures it, and its
simulation indicates that sharing environments would recover most of it at a
cost of milliseconds. Sharing needs the platform to change. Closing
each session when its conversation ends, rather than leaving it to the
platform's idle timeout, needs only the client.
Measuring the framework's CPU per step, the small number this price rests
on, took care: a stubbed model, a timer around the model call, or
steps run back to back each gave a plausible number that understated it. The open problem is for
platform builders: to confine each session as these runtimes do, without
reserving an environment for the time it spends waiting.

\textbf{Data availability.} The harness, deployment code, raw measurements,
the 3,549 deployed steps and the scripts for every figure and table are at
\url{https://github.com/MEKHALA/agent-runtime-capacity} (Apache 2.0).

\textbf{AI tool use.} The author used a generative AI tool, Claude
(Anthropic), to help write the text and the code, and is responsible for all
of the content.